\documentclass[reprint, amsmath,amssymb, aps, superscriptaddress]{revtex4-2}
\usepackage{amssymb}
\usepackage{amsmath}
\usepackage{epsfig}
\usepackage{color}
\usepackage{graphics, graphicx}
\usepackage{bbold}
\usepackage{psfrag}
\usepackage{mathcomp}
\usepackage{subfigure}
\usepackage{verbatim}
\usepackage{color}
\usepackage{float}
\usepackage{bm}
\usepackage[colorlinks, citecolor=blue]{hyperref}

\begin{document}
\title{Quantum-Fluctuation-Driven Dynamics of Droplet Splashing, Recoiling and Deposition in Ultracold Binary Bose Gases}
\author{Yinfeng Ma}
\affiliation{Beijing National Laboratory for Condensed Matter Physics, Institute of Physics, Chinese Academy of Sciences, Beijing 100190, China}
\author{Xiaoling Cui}
\email{xlcui@iphy.ac.cn}
\affiliation{Beijing National Laboratory for Condensed Matter Physics, Institute of Physics, Chinese Academy of Sciences, Beijing 100190, China}
\affiliation{Songshan Lake Materials Laboratory, Dongguan, Guangdong 523808, China}
\date{\today}
\begin{abstract}
Droplet impact on a surface is practically relevant to a variety of fields in nature and industry, while a complete control of its outcomes remains challenging due to various unmanageable factors. In this work, we propose the quantum simulation of droplet impact outcomes in the platform of ultracold atoms. Specifically, we study the quantum-fluctuation-driven dynamics (QFDD) of two-dimensional Bose-Bose mixtures from an initial Townes soliton towards the formation of a quantum droplet. By tuning the fluctuation energy of the initial Townes state through its size and number, the subsequent QFDD can produce various outcomes including splashing, recoiling, and deposition, similar to those in droplet impact dynamics. We have utilized the Weber number to identify the thresholds of splashing and recoiling, and further established a universal scaling law between the maximum spreading factor and the Weber number in the recoiling regime. In addition, we show that the residual QFDD in the deposition regime can be used to probe the collective breathing modes of a quantum droplet. Our results reveal a mechanism for the droplet impact outcomes, which can be directly tested in cold-atom experiments and can pave the way for exploring intriguing droplet dynamics in a clean and fully controlled quantum setting.  
\end{abstract}
\maketitle
\section{ Introduction}
Given the wide practical relevance to both nature and industry \citep{aerosol,erosion,ink1,coating2,cooling1,anti_icing,surface_material_review}, droplet impact dynamics on a surface has attracted much attention ever since the first study by Washington \cite{Worthington1,Worthington2}. Various impact outcomes-including splashing, receding/recoiling, rebound, and deposition-have been  observed successfully in experiments \citep{AS2001,PRL2005,JFM2006,PRL2012,ACIS2015,Langmuir2015,IJMF2017,PFluid2017,Langmuir2017a,Langumir2017b,ETFS2019,JCIS2019,PRL2019,ETFS2021,Langmuir2021,JFM2021}. In general, these dynamics were characterized by  two physical observables, namely, the maximum spreading factor ($\beta$) \citep{JFM2006,IJMF2017,ETFS2021} and the splashing threshold(${\cal K}$) \citep{PRL2005,PRL2012,ACIS2015,Langmuir2015,PFluid2017,Langmuir2017a,Langumir2017b,ETFS2019,JCIS2019,PRL2019,JFM2021},  which were shown not only to depend on the properties of the droplet itself (size, density, surface tension, viscosity, impact velocity), but also to be strongly influenced by the surface condition(roughness, wettability) and surrounding gas (pressure, composition).  Because of the complexities associated with various unmanageable factors, it is extremely challenging to deterministically parametrize $\beta,\ {\cal K}$ and fully control the impact outcomes. In this situation, a common practice is to assume an ideal droplet impact (on a smooth solid surface at atmospheric condition) and then to quantify the actual dynamics by the Weber and Reynolds numbers, which, respectively, describe the relative strength of droplet inertia with respect to capillary and viscous forces \citep{review1,review2}. Various scaling laws between $\beta,\ {\cal K}$ and these numbers have been proposed in the literature \citep{review1,review2}, based on different models or empirical fitting from experimental data. 

In the past few decades, ultracold atoms have emerged as an ideal platform for quantum simulation, given their extremely clean environment and the high controllability on the species, number, dimension, interaction strength etc \citep{RMP2008, RMP2010}. In particular, a recent important achievement in this field was the realization of a quantum droplet in both dipolar gas \citep{Pfau_1,Pfau_2,Pfau_3,Ferlaino,Modugno,Pfau_4,Ferlaino_2} and alkali bosonic mixtures \citep{Tarruell_1,Tarruell_2,Inguscio,Modugno_2,Modugno_3, DJWang}, with extremely dilute densities ($\sim 10^{14}- 10^{15} {\rm cm}^{-3}$) that can be eight orders of magnitude lower than water. In forming these gaseous droplets, quantum fluctuations play an essential role in providing the repulsive force for their stabilization, for which they are called quantum droplets \citep{Petrov}. To date, the idea of a quantum droplet has been successfully extended to various atomic systems, including low dimensional ones \citep{Petrov_2, Santos, Jachymski, Zin, Buchler, Cui3}, Bose-Fermi mixtures \citep{Cui2, Adhikari, Rakshit1, Rakshit2, Wenzel,Yi} and multi-component dipolar or alkali atomic mixtures \citep{Blakie,Santos_2,Ma}. The non-equilibrium properties of quantum droplets have also been investigated in terms of their dynamical formations \citep{dynamic1, dynamic2, dynamic3} and collisions \citep{collision1,collision2}. These developments offer an unprecedented opportunity for simulating droplet impact dynamics in ultracold atoms, particularly, at the microscopic quantum level and in a highly controllable manner.

In this work, we demonstrate the capability of using ultracold Bose gases to simulate the droplet impact outcomes in a fully controlled quantum setting. Contrary to the conventional droplet impact setup, here there is no impact surface for the droplet, and the driving force of its dynamics is purely from its intrinsic energy due to quantum fluctuations. Specifically, we study the dynamical property of a two-dimensional (2D) Bose-Bose mixture with repulsive intraspecies and attractive inter-species couplings, whose ground state is a quantum droplet. To highlight the quantum effect in the dynamics, we have chosen the initial state as the Townes soliton generalized from the single-species case \citep{Townes}, which features a zero mean-field energy with continuous scale invariance as recently confirmed in experiments \citep{Fibich,Hung1,Hung2,Beugnon}. In this way, the dynamics here is purely driven by quantum fluctuations and thus can be called quantum-fluctuation-driven dynamics (QFDD).  It is found that by tuning the fluctuation energy of the initial Townes state through its size($\sigma_0$) and number($N$), the subsequent QFDD can produce various outcomes, including splashing, recoiling and deposition, similar to those in droplet impact dynamics. We have mapped out the dynamical phase diagram in the ($\sigma_0,N$) plane and employed the Weber number to characterize different phases.  The splashing and recoiling thresholds are identified, and a universal scaling law is established between the maximum spreading factor and the Weber number for the recoiling dynamics, which is applicable for a considerably large parameter regime. Finally, we show that the long-time QFDD in the deposition regime can be used to extract the collective breathing modes of quantum droplet. These results can be directly tested in the current cold atoms experiments. In the Appendixes, we provide more details on the derivation of the
generalized Townes soliton with unequal masses, as well as on the numerical simulations.

\section{  Model}
 We start from the energy functional of two-species bosons in 2D ($\hbar=1$ for brevity):
\begin{eqnarray}\cal{E}(\bm{\rho})&=&-\sum_{i=1,2}  \phi_i^{*}(\bm{\rho}) \frac{\nabla_{\bm{\rho}}^2}{2m}\phi_i(\bm{\rho}) + \sum_{ij}\frac{g_{ij}}{2}n_i(\bm{\rho})n_j(\bm{\rho})\nonumber\\
&&+ {\cal E}_{\rm LHY}(n_i(\bm{\rho})); \label{E}
\end{eqnarray}
here $\bm{\bm{\rho}}=(x,y)$ is the 2D coordinate; $\phi_i$ is the wavefunction of the $i$-th species and $n_i=|\phi_i|^2$ is its density; $g_{ij}$ is the bare coupling between $i$- and $j$-species, which  can be expressed as $g_{ij}=4\pi a_{ij}/(ml_z)$ in quasi-2D geometry, with $a_{ij}$ the s-wave scattering length and $l_z$ the characteristic length along the confined ($z$) direction; and ${\cal E}_{\rm LHY}$ is the Lee-Huang-Yang(LHY) correction from quantum fluctuations, and for quasi-2D bosons near the mean-field instability point ($\delta a\equiv a_{12}+ \sqrt{a_{11}a_{22}}\sim 0$) it reads \citep{Petrov_2, Zin}
\begin{equation}
{\cal E}_{\rm LHY}=\frac{\eta^2 \ln(\eta\sqrt{e})}{8\pi m l_z^4},\ \ \ \ {\rm with}\ \  \eta=4\pi l_z(a_{11}n_1+a_{22}n_2).
\end{equation}
It has been shown that this LHY term can balance with the mean-field force and result in a self-bound droplet as the ground state \citep{Petrov_2}. Given the $n^2\ln(n)$ dependence of LHY energy, the 2D droplet can exist at both mean-field collapse and stable regimes \citep{Petrov_2} and with any infinitesimal atom number \citep{Cui3}. At a sufficiently large number, the 2D droplet develops a flat-top structure in its density profile, similar to the 3D case \citep{Petrov}.

Given Eq.(\ref{E}), the dynamics of  $\{\phi_i\}$ is governed by the time-dependent Gross-Pitaevskii(GP) equations:
\begin{equation}
i\partial_t\phi_i=\left(-\frac{\nabla_{\bm{\rho}}^2}{2m}+\sum_{j} g_{ij}n_{j}+\frac{\partial {\cal E}_{LHY}}{\partial n_i}\right)\phi_i. \label{GP}
\end{equation}
In this work, we will focus on the solution with zero angular momentum, since the ground state and the Townes soliton both stay within this sector, and different angular momentum sectors are decoupled from each other. In this sector, we can replace the coordinate $\bm{\bm{\rho}}$ simply by its magnitude $\rho=|\bm{\rho}|$. More details of solving Eq.(\ref{GP})  in the discretized coordinate and time space have been given in the Appendixes.

Throughout the paper, we specifically consider the two hyperfine states of $^{39}$K atoms, $|1\rangle\equiv|F=1,m_F=-1\rangle,\ |2\rangle\equiv|F=1,m_F=0\rangle$, as has been well studied in the droplet experiments \citep{Tarruell_1,Tarruell_2,Inguscio}. In this case,  $a_{11}=33.5a_B,\ a_{12}=-53a_B$ ($a_B$ is the Bohr radius), and $a_{22}$ is highly tunable by magnetic field. The confinement length is chosen as $l_z=0.08\mu m$. 

\section{ Generalized Townes soliton} For spinless bosons in 2D, it is known that the kinetic term and mean-field attraction can support a special stationary solution called the Townes soliton \citep{Townes}. This state features zero energy and continuous scale invariance, and it can only exist when the boson number and coupling strength satisfy $N|g|=5.85/m$. Such a special solution has been successfully observed in both non-linear optics \citep{Fibich} and ultracold atoms \citep{Hung1, Hung2, Beugnon}. In these experiments, the LHY correction takes little effect as it is much smaller than the mean-field part. In the following, we will show that the Townes soliton can be  generalized to two-species bosons if equally neglecting the LHY correction.

By omitting the LHY term in (\ref{GP}), we can see that the two GP equations for $\{\phi_1,\phi_2\}$ can support a single-mode solution $\phi_i=\sqrt{N_i}\phi \exp(-i\mu_i t)$ as long as 
\begin{equation}
\frac{N_1}{N_2} = \frac{g_{22}-g_{12}}{g_{11}-g_{12}}, \label{ratio}
\end{equation}   
where $\mu_1=\mu_2\equiv\mu$, and the single mode $\phi$ satisfies:
 \begin{equation}
\left(-\frac{\nabla_{\bm{\rho}}^2}{2m}+ Ng_{\rm eff}|\phi|^2\right)\phi=\mu\phi. \label{GP2}
\end{equation}
Here $N=N_1+N_2$ is the total number, and the effective interaction $g_{\rm eff}$ is given by
\begin{equation}
g_{\rm eff}=\frac{g_{11}g_{22}-g_{12}^2}{g_{11}+g_{22}-2g_{12}}. \label{geff}
\end{equation}
Apparently we have $g_{\rm eff}<0$ in the mean-field collapse regime ($g_{12}<-\sqrt{g_{11}g_{22}}$). It is then straightforward to check that under the condition 
\begin{equation}
Nm|g_{\rm eff}|=5.85,  \label{N_T}
\end{equation}
there exists a sequence of zero-energy eigenstates with continuous scale invariance, i.e., the eigenstate nature and zero-energy property will not change under an arbitrary scaling transformation $\phi(\bm{\rho})\rightarrow \lambda \phi(\lambda \bm{\rho})$ (accordingly $E\rightarrow\lambda^2E$). These stationary solutions are the generalized Townes soliton for two-species bosons. Note that a similar generalization also works for the case of unequal masses, with slight modifications in Eqs.(\ref{ratio},\ref{geff}) as shown in the Appendixes.
\begin{figure}
\centerline{\includegraphics[width=8.5cm]{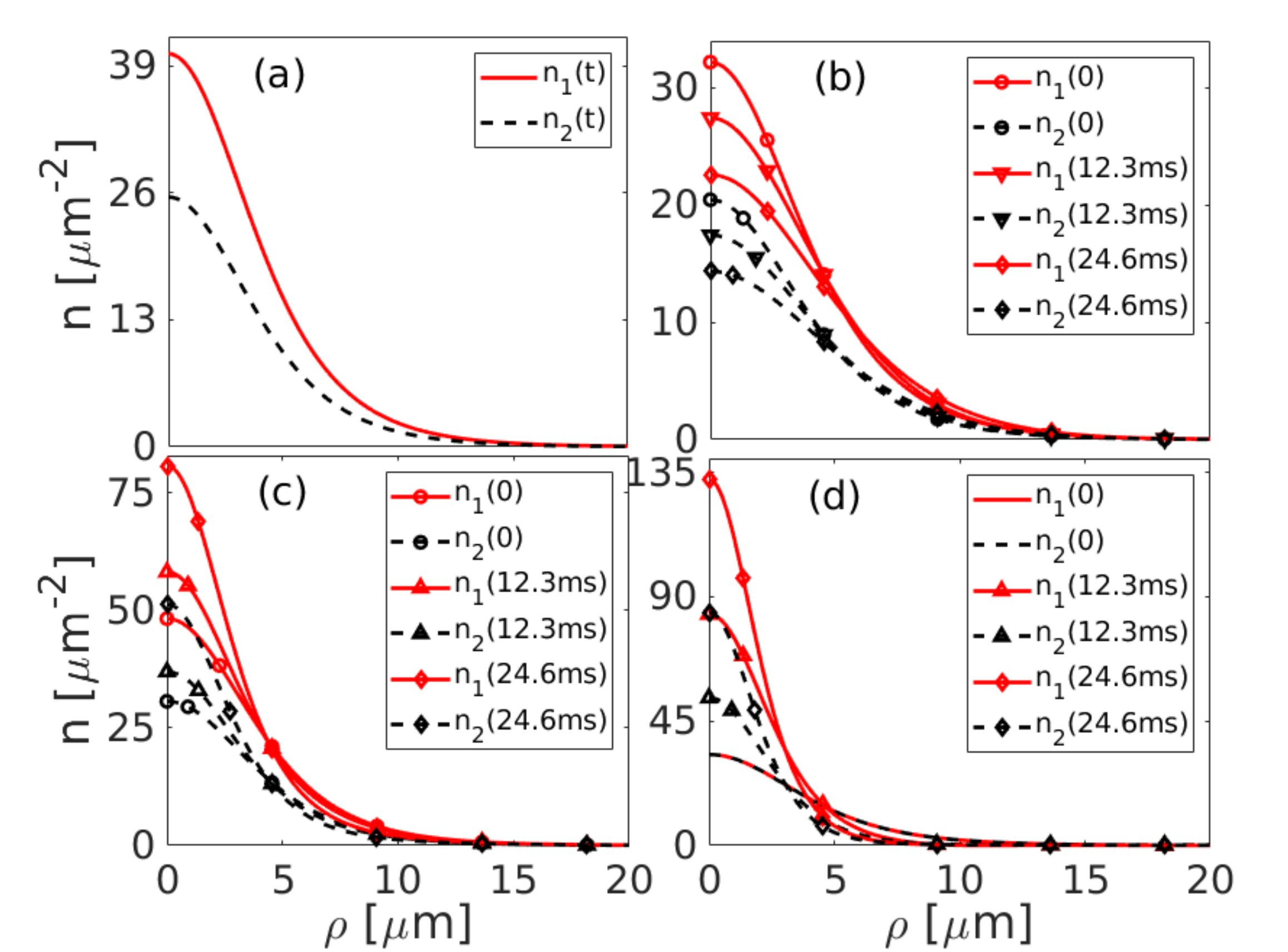}}
\caption{(Color online) Test of generalized Townes soliton for two-species bosons (neglecting the LHY correction). 
(a) A stationary Townes profile when the total number $N$ and number ratio $N_1/N_2$ satisfy (\ref{N_T}) and (\ref{ratio}) simultaneously. (b) Density profile at different times when the total number is changed to $0.8N$, while the number ratio is the same as in (a). (c) Density profile at different times when the total number is $1.2N$, with the same number ratio as in (a). (d) Density profile at different times with the same total number $N$ as in (a) while keeping $N_1/N_2=1$. Here we take two-species $^{39}$K atoms with $a_{11}=33.5a_B,\ a_{12}=-53a_B$ and $a_{22}=83.1a_B$, which gives $N=6500$ for a stationary Townes state according to Eqs.(\ref{geff},\ref{N_T});  the initial size of the Townes profile is taken as $7\mu m$.}
\label{fig_Townes}    
\end{figure}

In figure \ref{fig_Townes}(a), we confirm the stationary Townes soliton for two-species bosons once the total number $N$ satisfies (\ref{N_T}) and the number ratio $N_1/N_2$ satisfies (\ref{ratio}). In comparison, if we change $N$ to be smaller or larger, the original profile will shrink (figure \ref{fig_Townes}(b)) or expand (figure \ref{fig_Townes}(c)) with time. The profile is also unstable if $N_1/N_2$ deviates from (\ref{ratio}), see figure \ref{fig_Townes}(d). In a word, both conditions (\ref{N_T}) and (\ref{ratio}) are required in supporting a stationary two-species Townes solution.

\section{ Quantum-fluctuation driven dynamics} A crucial difference between the single- and two-species bosons is that quantum fluctuations play an important role in the latter, which can lead to droplet formation as a ground state. It then follows that starting from the generalized Townes soliton, which is  the mean-field stationary solution for two-species bosons, the quantum fluctuation can destabilize it strongly and drive its time evolution towards the droplet formation. Such dynamics can be called the quantum-fluctuation driven dynamics (QFDD), also in light of the fact that the total energy of Townes soliton is purely given by the LHY part, $E(t=0)=E_{\rm LHY}$.

\begin{figure*}
\centerline{\includegraphics[width=15cm]{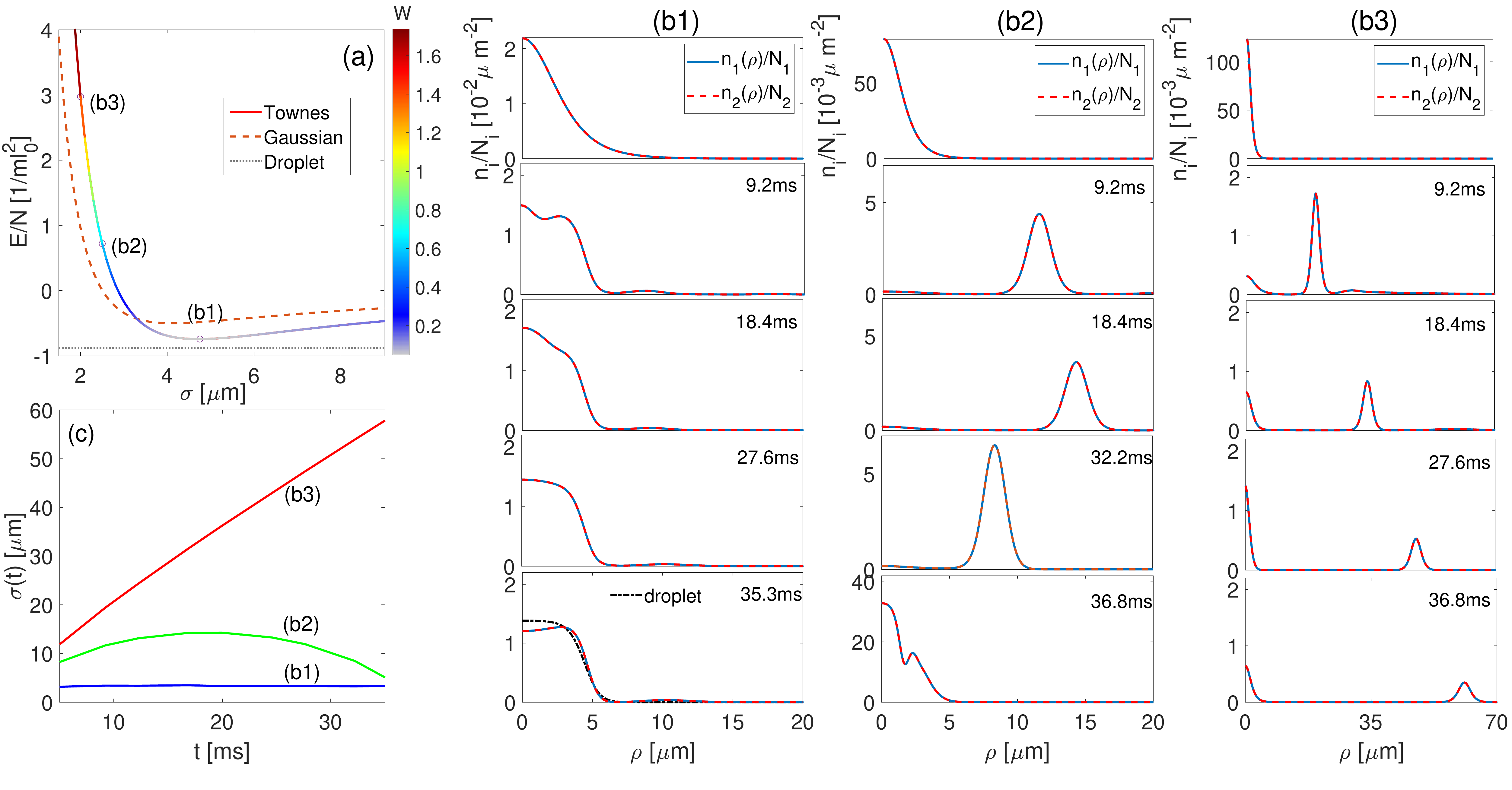}}
\caption{(Color online) Quantum-fluctuation driven dynamics(QFDD) of deposition (b1), recoiling (b2) and splashing (b3) starting  from initial Townes solitons with different sizes. The parameters of $\{a_{ij}\}$ and initial $N$ are the same as in figure \ref{fig_Townes}.     (a) Total energy per particle ($E/N=E_{\rm LHY}/N$) for the initial Townes state as functions of its  size ($\sigma$). The color on the curve denotes the corresponding Weber number. The dashed line shows the energy from Gaussian ansatz, and the lower dotted line shows the energy of ground state droplet. The energy unit is $1/(ml_0^2)$ with $l_0=1\mu m$.   (b1-b3) Three typical QFDD starting from different initial Townes states, which respectively simulate the droplet impact outcomes of deposition (b1), recoiling (b2) and splashing(b3). The initial sizes of Townes solitons are respectively $\sigma_0  (\mu m)=4.75(b1),\ 2.5(b2),\ 2(b3)$, as located by circles in (a). (c) Time evolution of the cloud size, $\sigma(t)$, for the dynamics in (b1-b3). }\label{fig_qfdd}
\end{figure*}

In figure \ref{fig_qfdd}(a), we show that the total energy ($=E_{\rm LHY}$) of two-species Townes soliton can be conveniently tuned by its size $\sigma$, taking a typical combination of $\{N, g_{ij}\}$ that satisfies (\ref{ratio},\ref{N_T}). In particular, $E_{\rm LHY}$ varies non-monotonically with $\sigma$ and shows a minimum at certain finite $\sigma$. To understand this behavior, we employ a Gaussian ansatz to approximate the single mode $\phi(\rho)=\exp[-\rho^2/(2\sigma^2)]/(\sqrt{\pi}\sigma)$, which leads to the total energy $E=\int 2\pi \rho d\rho {\cal E}(\rho)$ as
\begin{equation}
    \begin{split}
      E=\frac{N}{2m\sigma^2}+\frac{N^2g_{\rm eff}}{4\pi \sigma^2}+\frac{N^2\bar{a}^2}{ml_z^2\sigma^2}\ln \frac{4N\bar{a}l_z}{\sigma^2},
		\end{split}
	\end{equation} 
with $\bar{a}\equiv \sqrt{a_{11}a_{22}}$. We can see that the first two terms can support zero-energy states with arbitrary $\sigma$ under the condition $Nm|g_{\rm eff}|=2\pi$, which are just the simplified Gaussian version of the Townes soliton. However, when including the third LHY term, the total energy($=E_{\rm LHY}$) will deviate from zero, and the Townes profile is no longer stationary.  Given the expression $E_{\rm LHY} \sim -\ln(\sigma^2)/\sigma^2$, we can easily arrive at  a non-monotonic $E_{\rm LHY}\sim \sigma$ dependence with energy minimum $E_{\rm min}=-N\bar{a}/(4l_z^3e)$ at $\sigma_{\rm min}=\sqrt{4N\bar{a}l_z e}$. As shown in figure \ref{fig_qfdd}(a), $E_{\rm LHY}$ from a Gaussian ansatz provides a qualitatively good prediction to  the $E\sim\sigma$ lineshape of real Townes solutions.

Given the easily tunable fluctuation energy of the initial Townes state, the subsequent QFDD can exhibit rich dynamical outcomes. In figure \ref{fig_qfdd}(b1,b2,b3), we show the time evolution of density profiles for three typical QFDDs:

{\bf (I) Deposition.} When the initial size $\sigma_0$ is close to $\sigma_{\rm min}$ and $E_{\rm LHY}$ is small, the QFDD shows a typical deposition behavior (figure \ref{fig_qfdd}(b1)). Specifically, as time passes the system repels a small proportion of atoms outside, and the rest automatically follows the profile of a ground-state droplet with additional periodic oscillations. As discussed later, such residual oscillation can be used to probe the collective breathing modes of a quantum droplet. 

{\bf (II) Recoiling.} As $\sigma_0$ deviates more from $\sigma_{\rm min}$ and $E_{\rm LHY}$ gets larger, the system enters the recoiling regime. As shown in figure \ref{fig_qfdd}(b2), at early times a considerable portion of atoms are repelled outside, while at some point they stop to spread and flow back to merge with the central part. Such back-flow (or recoiling)  can be attributed to the competition between the surface tension and the kinetic energy of the cloud during the dynamics.

{\bf (III) Splashing.} When $\sigma_0$ deviates significantly from $\sigma_{\rm min}$ and $E_{\rm LHY}$ is large enough, the system shows a rapid splashing dynamics, see figure \ref{fig_qfdd}(b3). In this case, the large $E_{\rm LHY}$ causes a drastic change of the initial Townes profile in a short time, i.e., the cloud quickly splits into two pieces. In this process, $E_{\rm LHY}$ converts to the large kinetic energy of the outgoing part, such that it completely separates from the central part and flows away forever.

The above dynamics can be well distinguished by monitoring the mean size of the dynamical system, $\sigma(t)=\sqrt{\langle \bm{\rho}^2\rangle_t}$. As shown in figure \ref{fig_qfdd}(c), at longer times $\sigma(t)$ is almost static in the deposition regime, while it shows a non-monotonic behavior in the recoiling regime and a continuous increase in the splashing regime. 

To this end, we have shown that the QFDD starting from the Townes states can produce rich dynamical phases, including the deposition, recoiling, and splashing, which perfectly mimic the droplet impact outcomes as studied in the literature \citep{AS2001,PRL2005,JFM2006,PRL2012,ACIS2015,Langmuir2015,IJMF2017,PFluid2017,Langmuir2017a,Langumir2017b,ETFS2019,JCIS2019,PRL2019,ETFS2021,Langmuir2021,JFM2021}. However, different from these existing studies of droplet impact on a surface, in our case the driving force of the dynamics is purely from the intrinsic energy contributed  by quantum fluctuations. This introduces a mechanism for these fluid dynamics. Meanwhile, in QFDD there are no complexities caused by the impact surface or environment, and the dynamical outcome can be fully controlled by adjusting the size $(\sigma_0)$ and the number $(N)$ of the initial state.  Therefore, the ultracold atoms provide an extremely clean and convenient platform to simulate droplet dynamics, where the quantum effect can be well manipulated and the fluid mechanics can be understood in a more deterministic way. 

\section{Weber number and dynamical phase diagram} We now quantify various dynamical phases in QFDD by the Weber number. Note that the Reynolds number is irrelevant here because of the zero viscosity of Bose condensates. In conventional droplet impact dynamics \citep{review1,review2}, the Weber number is defined as $W=\rho_d D v^2/\gamma$, where $\rho_d,\ D, \ v,\ \gamma$ respectively denote the droplet density, diameter, impact velocity, and surface tension. It measures the relative strength of droplet inertia with respect to capillary force. Here, we generalize the  definition of $W$ to describe the quantum dynamics in general: 
\begin{equation}
W=\frac{\Delta E}{D \gamma}, \label{We}
\end{equation}
where $\Delta E$ is the energy difference between the initial state (here the Townes soliton)  and the true ground state given the same initial parameters ($N,\ g_{ij}$);  $D$ and $\gamma$ are the same as before, i.e., the droplet diameter and surface tension. Specifically, we have $\gamma=\int d\bm{\bm{\rho}}\left({\cal E}(\bm{\bm{\rho}})-\mu_1|\phi_1|^2-\mu_2|\phi_2|^2\right)$, with $\mu_i$ the chemical potential for the i-th species.

As shown in the color plot in figure \ref{fig_qfdd}(a), $W$ defined in (\ref{We}) can well characterize different dynamical phases in QFDD. Namely, a low $W$ corresponds to the deposition dynamics, where the small $\Delta E$ can be well absorbed by the surface change of the droplet;  as $W$ increases, the system enters the recoiling regime and finally end up at splashing, where the large $\Delta E$ overwhelms the capacity of the droplet surface and causes it to change drastically. Hereafter, we refer to the critical $W$ at the recoiling-splashing boundary as the splashing threshold (${\cal K}_s$), and that at the deposition-recoiling boundary as the recoiling threshold (${\cal K}_r$).

\begin{figure}[htbp]
\centerline{\includegraphics[width=7cm]{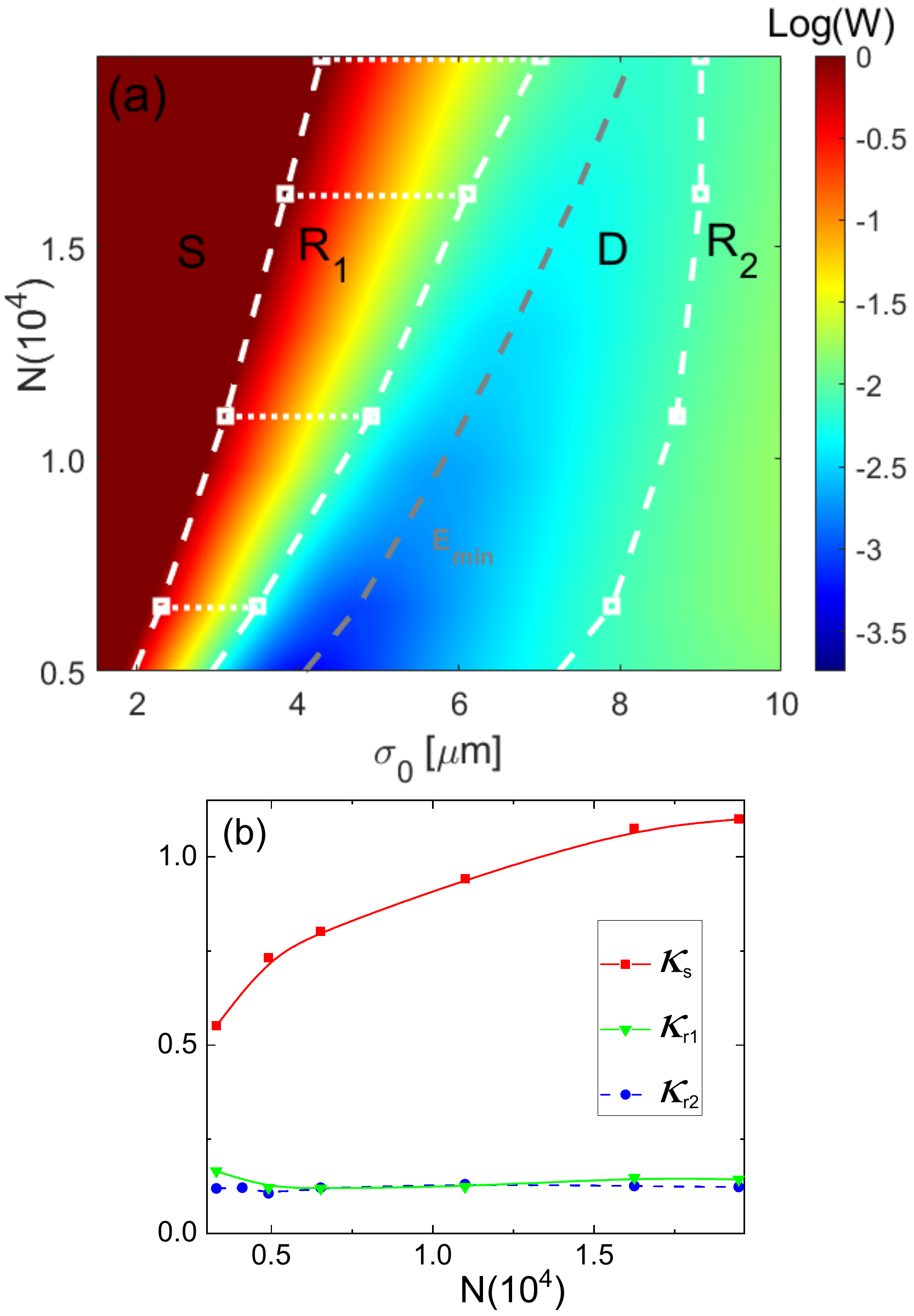}}
    \caption{(Color online) Characterization of different dynamical phases and their boundaries in-between. (a) Contour plot of Weber number in the parameter plane of initial Townes size ($\sigma_0$) and number($N$). The splashing, recoiling and deposition phases are respectively marked by 'S', 'R$_{1,2}$' and 'D' in the diagram. The white squares with white dashed lines mark their boundaries. Gray dashed line shows the location of energy minimum for initial Townes state. (b) Splashing threshold (${\cal K}_s$) and two recoiling thresholds (${\cal K}_{r1},{\cal K}_{r2}$) as functions of $N$ along the phase boundaries in (a). Note that when changing $N$, the coupling constant $a_{22}$ of $^{39}$K atoms also changes according to the  constraint (\ref{geff},\ref{N_T}) for initial Townes states. The other couplings $\{a_{11},a_{12}\}$ are the same as in figure \ref{fig_Townes}.}
\label{fig_diagram}
    \end{figure}

In figure \ref{fig_diagram}(a), we map out the dynamical phase diagram in the ($\sigma_0,N$) parameter plane. Distinct dynamical outcomes of QFDD, including deposition ('D'), recoiling ('R') and splashing ('S'), are identified by monitoring the mean size $\sigma(t)$ of the cloud during expansion (see figure \ref{fig_qfdd}(c)). In addition, we show the contour plot of $W$ in the ($\sigma_0,N$) plane, and one can see clearly that the 'D', 'R' and 'S' phases respectively correspond to the small, intermediate and large $W$ regions. Due to the non-monotonic dependence of $W$ on $\sigma_0$ (as shown in figure \ref{fig_qfdd}(a)), there are two recoiling regions in the diagram, as marked by 'R$_1$' and 'R$_2$'. In figure \ref{fig_diagram}(b), we extract the two recoiling thresholds(${\cal K}_{r1},{\cal K}_{r2}$) and the splashing threshold(${\cal K}_s$) along the phase boundaries as varying $N$. One can see that ${\cal K}_{r1}\approx{\cal K}_{r2}$ are given by a constant $\sim 0.12$, while ${\cal K}_s$  is a much larger value and continuously increases with $N$. 

\section{ Maximum spreading factor} Another important physical quantity to characterize the droplet impact dynamics is the maximum spreading factor $\beta=\sigma_{\rm max}/\sigma_0$, as defined by the ratio between the maximum spreading radius ($\sigma_{\rm max}$) and the initial one ($\sigma_0$). Clearly one has $\beta\sim 1$ for the deposition dynamics and $\beta\rightarrow \infty$ for splashing. An interesting behavior of $\beta$ shows up in the recoiling regime, where $\beta$ is finite and varies sensitively with $W$.

\begin{figure}[t]
\centerline{\includegraphics[width=8cm]{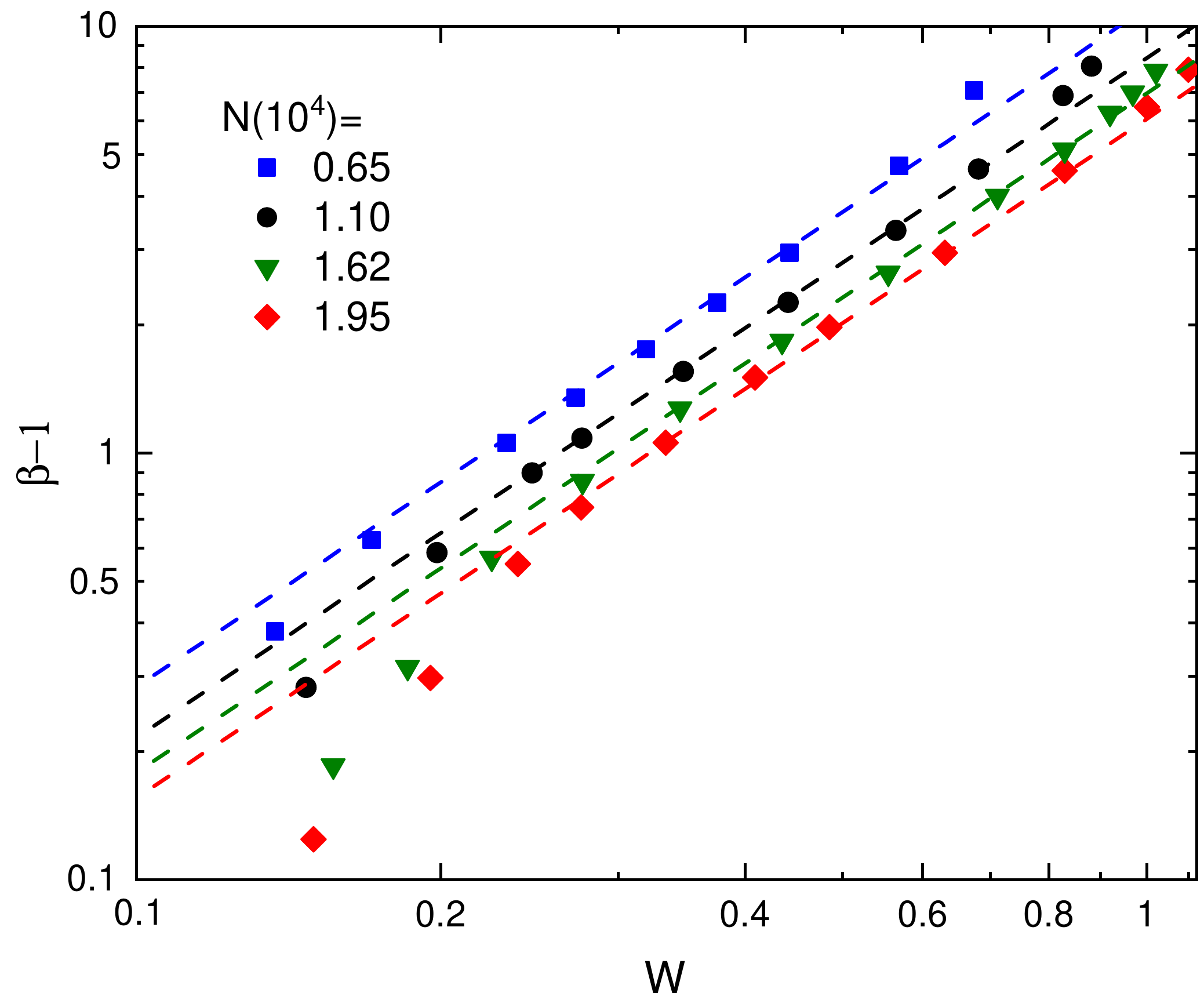}}
    \caption{(Color online) Universal scaling between the maximum spreading factor ($\beta$) and the  Weber number ($W$) in the recoiling regime. Here we take four given atom numbers of the initial Townes state with tunable sizes, as following the trajectory along the horizontal lines in Fig.\ref{fig_diagram}(a). The discrete points are from numerical calculations and the lines are fitting functions according to Eq.(\ref{fit}).} 
\label{fig_beta}
    \end{figure}

In figure \ref{fig_beta}, we extract $\beta$ as a function of $W$ along the horizontal lines in figure \ref{fig_diagram}(a), i.e., by varying $\sigma_0$ at several fixed $N$ in the recoiling regime. Apart from the region near the phase boundaries ($\beta\rightarrow 1$ or large enough), the data of $\beta$ for any given $N$ well follow the scaling relation:  
\begin{equation}
\beta-1 \propto W^{\alpha},\ \ \ {\rm with}\ \ \alpha=1.59. \label{fit}
\end{equation}
As shown in figure \ref{fig_beta}, such scaling works well for a wide parameter regime with $N/10^4\in (0.5,2)$, $\beta\in(1.5,10)$ and $W\in (0.2,1)$. Therefore, we expect the scaling law in (\ref{fit}) to reflect a very robust intrinsic property of the ultracold fluid during the recoiling QFDD.  

\section{ Breathing modes} Finally, we demonstrate that the deposition regime of QFDD can be used to probe the breathing modes of quantum droplet. The breathing mode can be theoretically obtained as follows. Assuming a small fluctuation mode $\delta\phi_i$ for the i-th species boson, and only keeping the lowest-order fluctuations in the GP equations(\ref{GP}), we obtain the following equations for $\{\delta\phi_i\}$:
\begin{widetext}  
\begin{equation}
\begin{split}
i\partial_t\delta\phi_1&=\left(-\frac{\nabla^2_{\bm{\bm{\rho}}}}{2m}+g_{11}n_1+g_{12}n_2+\frac{\partial {\cal E}_{LHY}}{\partial n_1}\right)\delta\phi_1+g_{11}n_1(\delta \phi_1+\delta\phi_1^*)+g_{12}\phi_1\phi_2(\delta \phi_2+\delta\phi_2^*)\\
&+\frac{\partial^2 {\cal E}_{LHY}}{\partial n_1^2}n_1(\delta \phi_1+\delta\phi_1^*)+\frac{\partial^2 {\cal E}_{LHY}}{\partial n_1\partial n_2}\phi_1\phi_2(\delta \phi_2+\delta\phi_2^*),  \\
\end{split}
\end{equation}
\begin{equation}
\begin{split}
 i\partial_t\delta\phi_2&=\left(-\frac{\nabla^2_{\bm{\bm{\rho}}}}{2m}+g_{12}n_1+g_{22}n_2+\frac{\partial {\cal E}_{LHY}}{\partial n_2}\right)\delta\phi_2+g_{22}n_2(\delta \phi_2+\delta\phi_2^*)+g_{12}\phi_1\phi_2(\delta \phi_1+\delta\phi_1^*)\\
&+\frac{\partial^2 {\cal E}_{LHY}}{\partial n_1\partial n_2}\phi_1\phi_2(\delta \phi_1+\delta\phi_1^*)+\frac{\partial^2 {\cal E}_{LHY}}{\partial n_2^2}n_2(\delta \phi_2+\delta\phi_2^*).  \\ 
\end{split}
\label{ceq}
\end{equation}
According to the standard Bogoliubov analysis, we search 
for solutions of the form:
\begin{equation}
  \delta \phi_i=\exp(-i\mu_it)\sum_j\left(u_{ij}(\bm{\rho})\exp(-i\omega_jt)+v_{ij}^*(\bm{\rho})\exp(i\omega_jt)\right).
\label{dpsi}
\end{equation}
Here $\omega_j$ is the $j$-th collective (eigen-)mode of the system. These modes can be extracted from the  following coupled equations for $u_{ij}(\bm{\rho})$ and $v_{ij}(\bm{\rho})$:
\begin{equation}
  \left(\begin{matrix}
    L_1+M_1&M_{12}&M_1&M_{12}\\
     M_{12}& L_2+M_2&M_{12}&M_2\\ 
   -M_1&-M_{12}&-(L_1+M_1)&-M_{12}\\
   -M_{12}&-M_2&-M_{12}&-(L_2+M_2)
  \end{matrix}
  \right)
   \left(\begin{matrix}
    u_{1j}\\
    u_{2j}\\
    v_{1j}\\
    v_{2j}
  \end{matrix}
  \right)
  =\omega_j\left(\begin{matrix}
   u_{1j}\\
    u_{2j}\\
    v_{1j}\\
    v_{2j}
  \end{matrix}
  \right),   \label{breathing}
  \end{equation}
\end{widetext}
where
\begin{equation}
\begin{split}
L_i&=-\frac{\nabla^2_{\bm{\bm{\rho}}}}{2m}+\sum_jg_{ij}|\phi_j|^2+\frac{\partial {\cal E}_{LHY}}{\partial n_i}-\mu_i\\
M_i&=g_{ii}|\phi_i|^2+\frac{\partial^2 {\cal E}_{LHY}}{\partial n_i^2}n_i\\
M_{12}&=g_{12}\phi_1\phi_2+\frac{\partial^2 {\cal E}_{LHY}}{\partial n_1\partial n_2}\phi_1\phi_2.\\
\end{split}
\end{equation}
In Fig.5, we show the lowest two breathing modes ($\omega_1,\omega_2$) below the atom emission threshold ($-\mu_1,-\mu_2$), see solid lines. In our numerics, we have obtained $\omega_j$ by discretizing the coordinate space and exactly diagonalizing the resulted large matrix  in the left-hand-side of (\ref{breathing}). Note that the breathing modes, by definition, stay in the same zero angular momentum sector as the Townes state and the ground state droplet.  More details on the numerics  are given in the Appendixes.

\begin{figure*}
 \centerline{\includegraphics[width=15cm]{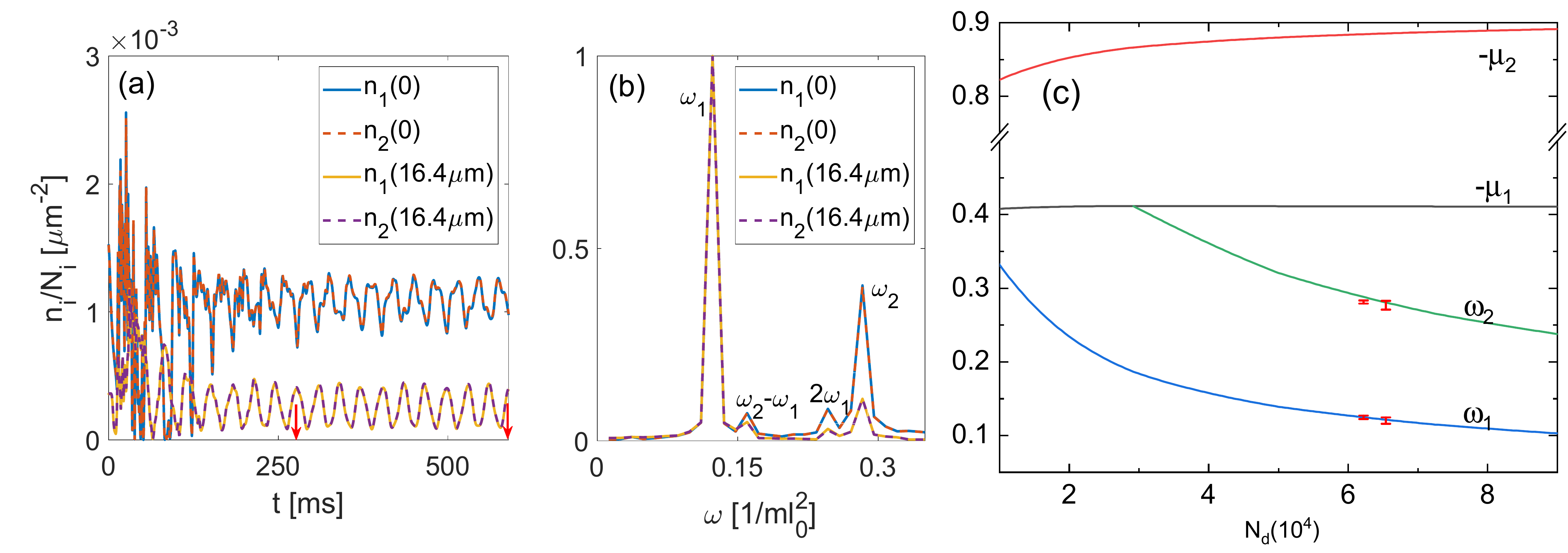}}
    \caption{(Color online) Extracting collective breathing modes from QFDD in the deposition regime. Here we take $a_{22}=83.8a_B$ and the other parameters are the same as in Figure \ref{fig_Townes}; the initial atom number is $N(10^4)=7.3$.  (a) and (b) show the boson densities at different locations  $\bm{\rho}=0$ (center), $16.4\mu m$ in the time and frequency domains (related by Fourier transformation), staring from an initial Townes state with size $\sigma_0=18\mu m$.  The two red arrows in (a) mark the time interval for Fourier transformation. 
(c) Collective excitation spectrum as a function of droplet number $N_d$. The lines are theoretical results based on Bogoliubov analysis, and the discrete data with error bar are extracted from the real density oscillations in the deposition dynamics. In the latter case the residual number $N_d$ can be effectively tuned by setting different sizes of initial Townes state, and here we take two sizes $\sigma_0(\mu m)=16,\ 18$ which lead to $N_d(10^4)=6.5,\ 6.2$ respectively. The energy unit is $1/(ml_0^2)$ with $l_0=1\mu m$. }
\label{fig_breathing}
 \end{figure*}

In figure \ref{fig_breathing}, we show how to extract collective breathing modes from the QFDD in the deposition regime for a given set of coupling strengths $\{a_{ij}\}$ and initial atom number $N$. Figure \ref{fig_breathing}(a) shows the typical density oscillations with time at  different locations. One can see that such oscillations are synchronic for both species and at  different locations, and therefore their Fourier transformations give the same peak frequencies; see figure \ref{fig_breathing}(b). In figure \ref{fig_breathing}(c), we plot out two sets of extracted peak frequencies ($\omega_1,\omega_2$) with different residual droplet numbers $N_d$. Here $N_d$ can be tuned effectively by setting different sizes of the initial Townes state. For all the extracted data we have collected 30 samples by varying the real-space locations or the time intervals in Fourier transformation, from which we obtain both the mean value and the variance.  One can see that the extracted results match very well with theoretical predictions of collective breathing modes from Eq.\ref{breathing} (solid lines). 

We note that a previous study has extracted one branch of a collective mode from the formation dynamics of a quantum droplet \citep{dynamic1}, while the dynamics there is not QFDD. Here, we have shown in figure \ref{fig_breathing} that two branches of collective modes can be simultaneously extracted from the residual oscillations in QFDD. Moreover,  we have checked that by taking different $\{a_{ij}\}$ and initial $N$, one can also get access to the other regime with only a single breathing mode. In comparison to the single-mode case, here the main complexity brought by the presence of two modes is  the appearance of additional frequency peaks at $\omega_2\pm\omega_1$ due to their interference. To correctly identify the breathing modes $\{\omega_j\}$ from the multiple frequency peaks, it is important to note that these modes are associated with the most pronounced peaks among all others.  This can be understood by analyzing the spin densities $n_i=|\phi_i+\delta\phi_i|^2$, with $\delta\phi_i$ given in Eq.(\ref{dpsi}). Obviously, one can see that the Fourier transform of $n_i$ results in multiple peaks at $\{\omega_j\}$, $\{2\omega_j\}$ and $\{\omega_i\pm\omega_j\}$, while the ones at $\{\omega_j\}$ have the largest magnitudes, which depend linearly on the strength of the fluctuation mode (the others all have a quadratic dependence). Based on this principle, one can easily identify the correct breathing modes ($\omega_1,\omega_2$) as marked in figure \ref{fig_breathing}(b). 

\section{ Experimental relevance}
Our results can be readily tested in the current cold-atom experiments. The initial Townes profile with a given amplitude and size can be imprinted by properly designing the optical potential applied to the atoms, as successfully implemented in previous experiments on two-component Bose gases \citep{Beugnon,DMD}. The subsequent dynamics of the system can then be measured through the in situ image, and various dynamical outcomes can be distinguished typically within tends of milliseconds, as shown in Fig.\ref{fig_qfdd}. Within this time scale severe atom loss in the $^{39}$K droplet can be effectively avoided \citep{Tarruell_1}. On the other hand, for long time dynamics, the atom loss will play an essential role and may even impede the breathing mode detection, which requires a typical time scale of hundreds of milliseconds, as shown in Fig.\ref{fig_breathing}(a1,a2). For that, one can resort to the heteronuclear mixtures of $^{41}$K and $^{87}$Rb, whose lifetime has been shown to be much longer ($\sim 1s$)  due to the low droplet density therein to suppress the atom loss \citep{Modugno_2, Modugno_3}. Since the Townes soliton can be well extended to boson mixtures with unequal masses (see the Appendixes), and the controllability  of LHY energy by the size and number generally applies to these systems, we expect that various dynamical outcomes and the breathing mode extraction can be tested equally in heteronuclear Bose gases.

\section{ Summary and discussion} In summary, we have demonstrated the quantum simulation of droplet impact outcomes in ultracold boson mixtures. Various dynamical phases including splashing, recoiling, and deposition, have been revealed. A remarkable difference here is that these dynamics are purely driven by quantum fluctuations, instead of the mechanical impact force in previous studies. Given the easy manipulation of initial states with tunable fluctuation energies, the current cold-atom platform provides complete control of these dynamics in the microscopic quantum level. To characterize different dynamics, we have introduced the Weber number and examined two important physical quantities, namely, the splashing/recoiling threshold and the maximum spreading factor. We have also proposed to extract the collective breathing modes from the residual dynamics in the deposition regime. These results are directly relevant to ongoing cold-atom experiments.

Furthermore, we remark that our work is in distinct contrast to the previous studies of QFDD in ultracold atoms, where a visible dynamics can only be achieved for small condensates \citep{Law, Cui, Barnett, Heinze}. This is because usually the energy difference per particle between the mean-field and the true quantum ground states decays rapidly as the number $N$ increases \citep{Ueda, Ho2, Muller, Zhou} , and thus for large $N$ the quantum fluctuation energy is negligibly small as compared to the mean-field one. However, this is not the case for a quantum droplet. For a static droplet, quantum fluctuations have been shown to provide an indispensable repulsive force for its stabilization, without which the whole system will collapse \citep{Petrov}. In parallel, our current work reveals the equally significant role of quantum fluctuations played in the non-equilibrium dynamics of a quantum dropet, regardless of whether the droplet is small or large. 

Finally, we anticipate that the dynamical outcomes of splashing, recoiling, and deposition revealed in this work may be equivalent with other initial conditions when the dynamics is not purely driven by quantum fluctuations. In fact, the generalized Weber number as defined in Eq.(\ref{We}) can be applied to any initial state, and the key issue is whether the additional energy of such a state (as compared to the true ground state) can be absorbed by the surface tension of a droplet during dynamics. In this sense, Eq.(\ref{We}) may serve as a unified quantity to understand the different dynamics of quantum droplets with various initial conditions.

\section*{ Acknowledgment}
The work is supported by the National Key Research and Development Program of China (2018YFA0307600), the National Natural Science Foundation of China  (12074419, 12134015, 12205365), and the Strategic Priority Research Program of Chinese Academy of Sciences (No. XDB33000000).

\appendix

\section{Townes soliton for two-species bosons with unequal masses}

Neglecting the LHY correction, we write down the Gross-Pitaevskii(GP) equation for two-species bosons with unequal mass ($m_1,m_2$):
\begin{align}
&i\partial_t \phi_1=\left(-\frac{\nabla_{\bm{\rho}}^2}{2m_1}+g_{11}|\phi_1|^2+g_{12}|\phi_{2}|^2\right)\phi_1;\\
&i\partial_t \phi_2=\left(-\frac{\nabla_{\bm{\rho}}^2}{2m_2}+g_{12}|\phi_1|^2+g_{22}|\phi_{2}|^2\right)\phi_2.
\end{align}
In the case $m_1\neq m_2$, we define the mass-imbalance parameter as $w=m_2/m_1$.
By comparing the two GP equations, we can see that they can support a single-mode solution $\phi_i=\sqrt{N_i}\phi \exp(-i\mu_i t)$ as long as 
\begin{equation}
\frac{N_1}{N_2} = \frac{wg_{22}-g_{12}}{g_{11}-wg_{12}}, \label{ratio2}
\end{equation}   
 where $\mu_1=w\mu_2$, and the single mode $\phi$ satisfies:
 \begin{equation}
\left(-\frac{\nabla_{\bm{\rho}}^2}{2m_1}+ Ng_{\rm eff}|\phi|^2\right)\phi=\mu_1\phi. \label{GP3}
\end{equation}
Here $N=N_1+N_2$ is the total number, and the effective interaction $g_{\rm eff}$ is given by
\begin{equation}
g_{\rm eff}=\frac{w(g_{11}g_{22}-g_{12}^2)}{g_{11}+wg_{22}-(w+1)g_{12}}. \label{geff2}
\end{equation}
Again we have $g_{\rm eff}<0$ in the mean-field collapse regime ($g_{12}<-\sqrt{g_{11}g_{22}}$), and the Townes solution occurs when 
\begin{equation}
Nm_1|g_{\rm eff}|=5.85.  \label{N_T2}
\end{equation}
For the equal mass case ($w=1$), these equations automatically reduce to Eqs.(4-7) in the main text.

\section{Numerical Method}

In our numerical simulations, we have considered the solution with zero angular momentum. This choice is because the angular momentum is preserved by the Hamiltonian, and both the ground state and the Townes soliton stay exactly in the zero angular momentum sector. Moreover, the breathing mode, by definition, also stays in the zero angular momentum sector. Given this constraint, in the following we simply replace the coordinate $\bm{\bm{\rho}}$ by its magnitude $\bm{\rho}=|\bm{\bm{\rho}}|$.

In solving the GP equation of the form $i\partial_t \phi(\rho,t)=H\phi(\rho,t)$, we have discretized the coordinate and time as $\rho_j=j\delta, \ t_n=n\tau$, where $j,n$ are all integers and we have taken small intervals $\delta=0.005\mu m,\ \tau=2.74\times 10^{-3}ms$. Then starting from a given state $\phi_n$ at time $t_n$, we obtained the new  wavefunction $\phi_{n+1}$ at the next time $t_{n+1}$ as follows. 

First, we split $H$  into the local and non-local parts: $H=H_1+H_2$, where the derivatives of $\rho$ are all contained in  $H_2$. In this case, we simply have $H_2$ as the kinetic energy $H_2=-\frac{1}{2m}\left[\frac{\partial^2}{\partial\rho^2}+\frac{1}{\rho}\frac{\partial}{\partial \rho}\right]$, and $H_1$ is the rest part that solely depends on local densities. 

As the first step of the update, the local $H_1$ produces an  intermediate state $\phi'$ from $\phi_n$:
\begin{equation}
\phi'=\exp(-iH_1\tau)\phi_n.
\label{s24}
\end{equation}
Then, we perform the time evolution generated by $H_2$ with the Crank-Nichilson scheme:
\begin{equation}
\frac{\phi_{n+1}-\phi'}{-i\tau}=\frac{H_2}{2}\left(\phi_{n+1}+\phi'\right),
\end{equation}
which gives 
\begin{equation}
\phi_{n+1}=\frac{1-i\tau H_2/2}{1+i\tau H_2/2}\phi'.
\label{s27}
\end{equation}
 Specifically, in the discretized coordinate space we have 
\begin{equation}
H_2\phi_j=-\frac{1}{2m}\left(\frac{\phi_{j+1}+\phi_{j-1}-2\phi_j}{\delta^2}+\frac{1}{j\delta}\frac{\phi_{j+1}-\phi_{j-1}}{2\delta}\right).
\end{equation}
Here we have simplified $\phi(\rho_j)$ as $\phi_j$. To this end, (\ref{s27}) gives the updated wavefunction $\phi_{n+1}$ at time $t_{n+1}$.

In above we have shown the details on the numerical simulation of GP equations. By transforming $\tau\to-i \tau$, i.e., the imaginary time evolution, one can also obtain the ground state of the system (droplet solution). To  make sure that the wavefunction is reduced to zero well before touching the boundary of the system,  we have chosen a sufficiently large size in the simulation with the maximum radius ranging from 50$\mu m$ to 200$\mu m$. In this way one can avoid the boundary effect to dynamical and static properties of the system.

In solving the breathing modes from Eq.(\ref{breathing}) in the main text, we have again discretized the coordinate space and transformed it to a large matrix equation.  The breathing modes are then obtained by exactly diagonalizing the matrix.

\end{document}